\documentclass[preprint,aps,amsmath,superscriptaddress,nofootinbib,tightenlines,showpacs]{revtex4}
\usepackage{amsmath}
\usepackage{amsfonts}
\usepackage{amssymb}
\usepackage{latexsym}
\usepackage{graphicx}
\usepackage{bm}
\usepackage{epsfig}
\usepackage[english]{babel}
\usepackage[bookmarks,colorlinks,breaklinks,hypertex]{hyperref}

\def\c{\mathrm{c}}
\def\eV{~\mathrm{eV}}
\def\keV{~\mathrm{keV}}
\def\MeV{~\mathrm{MeV}}
\def\GeV{~\mathrm{GeV}}

\begin{document}


\title{Mass-dependent Lorentz Violation and Neutrino Velocity}
\author{Miao Li\footnote{Electronic address: mli@itp.ac.cn}}
\affiliation{Institute of Theoretical Physics, Chinese Academy of Sciences,\\
Beijing 100080, China\\}
\affiliation{Kavli Institute for Theoretical Physics China, Chinese Academy of Sciences,\\
Beijing 100080, China\\}
\affiliation{State Key Laboratory of Frontiers in Theoretical Physics, Chinese Academy of Sciences,\\
Beijing 100190, China\\}
\author{Tower Wang\footnote{Electronic address: twang@phy.ecnu.edu.cn}}
\affiliation{Department of Physics, East China Normal University,\\
Shanghai 200241, China\\}
\affiliation{Kavli Institute for Theoretical Physics China, Chinese Academy of Sciences,\\
Beijing 100080, China\\} \vspace{0.2cm}
\begin{abstract}
Motivated by a recent and several earlier measurement results of the
neutrino velocity, we attempt to resolve the apparent discrepancies
between them from the viewpoint of mass-energy relation in special
relativity. It is argued that a complicated tachyonic neutrino model
or a mass-dependent Lorentz violation theory can do this job.
\end{abstract}


\maketitle



\section{Motivation}\label{sec:mot}
Over the past one hundred years, special relativity has been one
cornerstone of modern physics, well-established by innumerable
experiments and observations. An outstanding feature of special
relativity is a universal upper limit of speed, namely the light
speed $c$ in vacuum. Astonishingly, according to a recent result
\cite{OPERA:2011zb}, this speed record is broken by neutrinos in the
OPERA experiment, confirming an earlier record from the MINOS
detector \cite{Adamson:2007zzb}.

Before drawing a conclusion, certainly it is most imperative to
check further systematic errors in the measurement. But it is also
intriguing to tentatively make a concession to experimental data and
adjust our theories. This paper is a preliminary attempt to resolve
the neutrino velocity anomaly with a complicated tachyonic neutrino
model and with a tiny deformation of Einstein's mass-energy
relation.

The paper is organized as follows. In Sec. \ref{sec:record} we
summarize main results of the neutrino velocity measurements from
literature and remark on some interesting features. In Sec.
\ref{sec:tach}, we attempt to explain the velocity data in framework
of the standard mass-energy relation by introducing a tachyonic mass
to neutrinos. We find the tachyonic neutrino model cannot reconcile
all data unless the neutrino mass runs with energy. Sec.
\ref{sec:lv} deviates from the standard mass-energy relation and
shows how the wired neutrino velocity can be explained by a scenario
of mass-dependent Lorentz violation. Models of this scenario are
classified in two categories: democratic models universal for all
types of matter, and discriminative models sensitive to particle
species. Some open questions are covered in Sec. \ref{sec:disc}.

\section{Neutrino Speed Records}\label{sec:record}
According to \cite{OPERA:2011zb} and references therein, hitherto
there are four independent measurements accomplished for neutrino
velocity. We summarize their results in Table \ref{tab-nuspeed}. In
the table we include neutrino flavors, which is helpful for
flavor-dependent explanations of these data.

Needless to say, a superluminal velocity of neutrino is highlighted
by the positive results $(v-c)/c>0$ from recent MINOS and OPERA
measurements. It conflicts apparently with basic laws in special
relativity. On the other hand, the old Fermilab and SN 1987A results
are consistent with special relativity. This is easy to check with
Eq. \eqref{m-E} below, provided the neutrino mass is of order
$10^{-1}\eV/c^2$. However, since Fermilab and SN 1987A did not tell
us the signature of $(v-c)/c>0$, their results still admit a
superluminal velocity of neutrino.

The SN 1987A data is unique. Some comments are in order on this
point. First, its source and baseline are different from others. The
neutrinos have traveled for about 168 thousand years before
detection. Second, in contrast to roughly ten thousands of events in
other experiments, only a dozen of events are detected for SN 1987A
burst. This leads to a large statistic error not shown in Table
\ref{tab-nuspeed}. Third, thanks to the astronomical length scale of
``baseline'', SN 1987A puts a very stringent limit on neutrino
velocity, which is violated by both MINOS and OPERA experiments.
Each single trustable event from SN 1987A could put the MINOS and
OPERA results in doubt. Reverse the logic, were
$(v-c)/c\sim+2\times10^{-5}$ true for $\MeV$ neutrinos, these
neutrinos should have advanced their arrival for years than gamma
rays from SN 1987A. The lesson is, when formulating theories, one
must pay special attention to the SN 1987A constraint.

The results summarized in Table \ref{tab-nuspeed} exhibit dependence
on energy, flavor and time/century/baseline. Such features are
illuminating. The purpose of this paper is to explore the energy
dependence from the viewpoint of mass-energy relation and Lorentz
symmetry. We will adhere to Einstein's mass-energy relation in Sec.
\ref{sec:tach} and deviate from it slightly in Sec. \ref{sec:lv}.
\begin{table*}
\caption{Summary of neutrino velocity measurements. $v$ is the
measured average speed of neutrino and $c$ is the velocity constant
of light. The neutrino flavors are mostly identified at sources, but
for SN 1987A it is chosen for detectors. Limited by space, we have
not given the number of events in each experiment. See references
for details. \label{tab-nuspeed}}
\begin{tabular}{|c|c|c|c|c|c|}
\hline \hline
Experiment & Velocity constraint & Energy range & Flavors & Reference\\
\hline
Fermilab  & $|v-c|/c<4\times10^{-5}$ (95\% CL) & $30$ to $200\GeV$ & $\pi/$K-decay $\nu,\bar{\nu}$& \cite{Kalbfleisch:1979rm,Alspector:1976kd}\\
\hline
SN 1987A & $|v-c|/c<2\times10^{-9}$ & $5$ to $40\MeV$ & $\bar{\nu}_{e}$ and $\bar{\nu}_{\mu},\bar{\nu}_{\tau}$ & \cite{Longo:1987ub,Hirata:1987hu,Bionta:1987qt}\\
\hline
MINOS & $(v-c)/c=(5.1\pm2.9)\times10^{-5}$ (68\% CL) & $\sim3\GeV$ & $\nu_{\mu}$ and $\bar{\nu}_{\mu},\nu_{e},\bar{\nu}_{e}$ & \cite{Adamson:2007zzb}\\
\hline
OPERA & $(v-c)/c=(2.48\pm0.28\pm0.30)\times10^{-5}$ (6.0$\sigma$) & $\sim17\GeV$ & $\nu_{\mu}$ and $\bar{\nu}_{\mu},\nu_{e},\bar{\nu}_{e}$ & \cite{OPERA:2011zb}\\
\hline \hline
\end{tabular}
\end{table*}

\section{Troubles for Tachyonic Neutrino Models}\label{sec:tach}
A superluminal velocity can be easily achieved by tachyon with a
negative value of squared mass. At the same time, the neutrino mass
has been an enigma for half a century. Neutrino oscillation
experiments measure the difference of squared masses between
orthogonal mass eigenstates. Therefore, it sounds natural to
consider neutrinos as tachyons in light of the superluminal
phenomenon\footnote{The hypothesis of tachyonic neutrinos was
previously employed to explain certain anomalies in $\beta$-decay
experiments \cite{Ciborowski:1996qg}, but the ``mass'' scale there
is much lower than those in this section.}. In this section, we will
first develop the simplest version of this idea and then rule it
out. After that, we will pursue the possibility of a tachyonic
neutrino mass running with energy.

For convenience, we write down Einstein's mass-energy relation in
the form
\begin{equation}\label{m-E}
1-\frac{v^2}{c^2}=\frac{m^2c^4}{E^2},
\end{equation}
which relates energy $E$ and rest mass $m$ of a particle via its
velocity $v$ and the light velocity constant $c$. For ordinary
particles, nonnegativity of the right hand side dictates on the left
hand side $v\leq c$ as a well-known feature of special relativity.
Inversely, violation of this bound can be realized if we extrapolate
relation \eqref{m-E} to tachyons with $m^2<0$. For small deviation
$|v-c|/c\ll1$, \eqref{m-E} reduces to $2(v-c)/c\sim-m^2c^4/E^2$.
Applied to data, it gives $-m^2\sim(120\MeV/c^2)^2$ for OPERA,
$-m^2\sim(30\MeV/c^2)^2$ for MINOS, $|m^2|<(0.32\keV/c^2)^2$ for SN
1987A and $|m^2|<(0.27\GeV/c^2)^2$ for Fermilab. In the simplest
tachyon model, a constant tachyonic mass of neutrino cannot explain
all of the results in Table \ref{tab-nuspeed}. Within this model,
the OPERA and MINOS results are marginally at odds with each other,
and both of them conflict obviously with the SN 1987A bound.

But one may still imagine that, instead of a constant parameter,
$m^2$ is running with energy $E$. Because of an unknown mechanism,
its running is much larger than the prediction of standard model of
particle physics. Then one may reconcile the discrepancies mentioned
above by tuning the dependence of $m^2$ on $E$. Specifically, one
might be able to tune the running so that $dm^2/d\ln E<0$ for
$E<E_{\c}$ and $dm^2/d\ln E>0$ for $E>E_{\c}$. This possibility
overlaps partly the discriminative models of mass-dependent Lorentz
violation in Sec. \ref{subsec:disc}. An illustrative example and
more discussion will be given there.

We warn that tachyonic modes, like ghosts, generally give rise to
instabilities in quantum field theory. A tachyonic neutrino model
suffers from the same problem, no matter $m^2$ is a constant or not.
Therefore, it is challenging to construct viable models of this
type.

\section{Mass-dependent Lorentz Violation}\label{sec:lv}
As we have seen in the previous section, a tachyonic mass of
neutrino cannot explain observational data in a simple way. So let
us try another scenario by deviating from the Lorentz symmetry
slightly. Before doing this, we remind the readers that Lorentz
invariance is a symmetry of Minkowski metric and that Lorentz boosts
are related to some spaceime-dependent conserved quantities.
Theoretically, a deformation of Lorentz symmetry can be realized by
``curving'' the Minkowski spacetime. Observationally, the violation
of several conservation laws sheds light on testing our scenario in
other approaches.

The main insight is that results in Table \ref{tab-nuspeed} can be
accommodated with a deformation of mass-energy relation \eqref{m-E}
as
\begin{equation}\label{dm-E}
1-\frac{v^2}{c^2}=\lambda-f(\lambda),
\end{equation}
where $\lambda=m^2c^4/E^2$ and $f(\lambda)$ is an unknown function
with the property
\begin{equation}
f(\lambda)=\left\{\begin{array}{ll}
+\mathcal{O}(10^{-5}),~~~~&\mathrm{for}~\lambda~\mathrm{around}~\lambda_{\c}\sim\frac{m_\nu^2c^4}{(10\GeV)^2},\\
0,~~~~&\mathrm{elsewhere}.
\end{array}\right.
\end{equation}
The deformed relation has a physical interpretation as follows. For
any particle of mass $m_i$, there is a critical energy scale
\begin{equation}
E_{\c i}=\frac{m_ic^2}{\sqrt{\lambda_{\c}}}
\end{equation}
around which the Lorentz symmetry is broken down. The amount of
deviation from Lorentz symmetry is determined by $f(\lambda)$. In
viewpoint of the standard mass-energy relation, the particle gains a
tachyonic effective mass near its critical energy scale. Compared
with other scenarios of Lorentz violation, a new feature here is the
dependence on mass-energy ratio $\sqrt{\lambda}$. In other words,
the Lorentz violation is mass-dependent in this scenario. More
interestingly, for massless photons, relation \eqref{dm-E} reduces
to the standard mass-energy relation if we promote velocity to
momentum via $v/c=p/E$.

Based upon the nature of function $f(\lambda)$, this scenario can be
divided into two branches. In the first branch, $f(\lambda)$ is a
universal function for all species of matter. Both the form and
parameters of this function are universal. Then $\lambda_{\c}$ is a
universal constant. We will use ``democratic models'' to name models
in this branch and study them in Sec. \ref{subsec:demo}. Models in
the second branch will be dubbed ``discriminative models''. In these
models, the exact form and parameters of function $f(\lambda)$
depend on species of matter. For instance, the function $f(\lambda)$
could involve charges and coupling constants of the particle. The
discriminative models are investigated in Sec. \ref{subsec:disc}.

\subsection{Democratic Models}\label{subsec:demo}
Most economically $f(\lambda)$ is a universal function for all
species of matter irrespective of masses, flavors, charges,
\emph{etc}. Then $\lambda_{\c}$ is approximated by a universal
constant. Assuming $m_\nu\sim10^{-1}\eV/c^2$ and the critical energy
scale $E_{\c\nu}\sim10\GeV$ for neutrino, we get
$\sqrt{\lambda_{\c}}\sim10^{-11}$. For electron of mass
$0.5\MeV/c^2$, this yields a critical energy $E_{\c e}\sim10^7\GeV$,
a scale well above the CERN LHC energy. For heavier particles, the
critical scale is even higher. In this perspective, the mass
hierarchy of standard model particles is responsible for the energy
hierarchy of Lorentz-violating new physics. And quite fortunately,
accelerator-generated neutrinos fall in such a weird ``new physics''
scale. The hierarchy also puts our scenario in safety. So far we
have failed to find a counter example for this scenario. But it is
believed that probably the scenario of this branch can be ruled out
by a clever reasoning or by some data in the large body of
experiments.

\begin{figure}
\begin{center}
\includegraphics[width=0.45\textwidth]{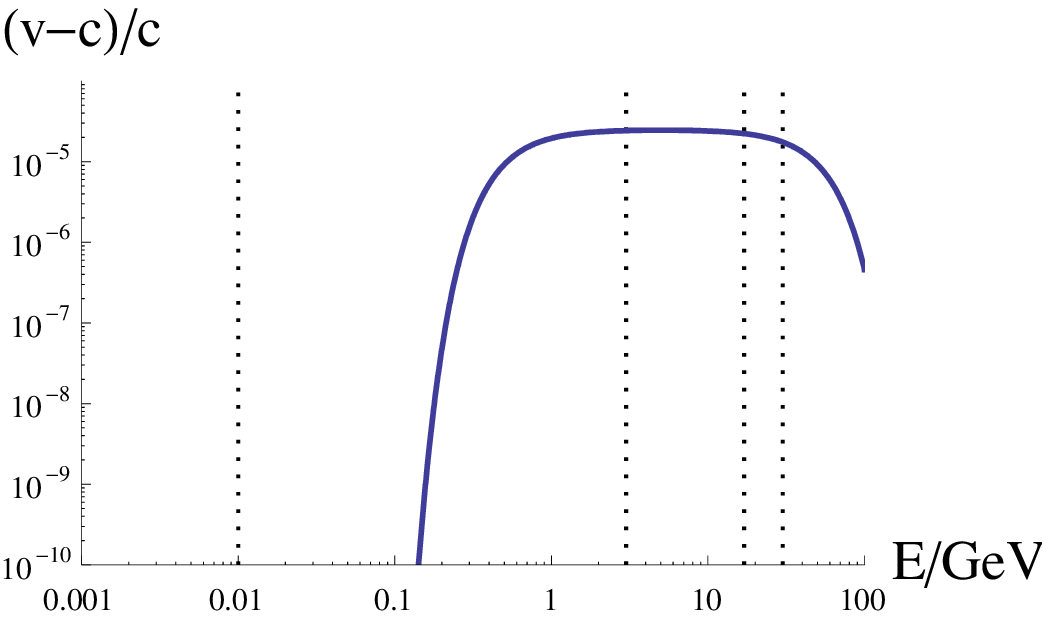}~~
\includegraphics[width=0.45\textwidth]{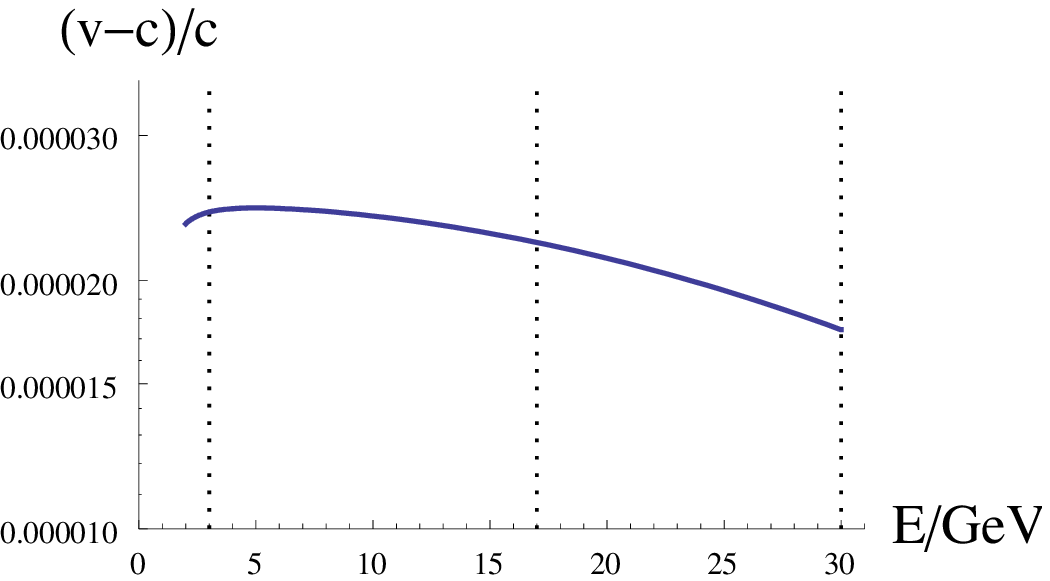}\\
\end{center}
\caption{(color online). A numeric illustration of the deformed
mass-energy relation \eqref{dm-E} with $f(\lambda)$ given by the toy
model \eqref{toy}. The model parameters are chosen as
$m_{\nu}=10^{-1}\eV/c^2$, $\epsilon=0.01$, $\delta=5\times10^{-5}$
and $E_{\c\nu}=5\GeV$. It accommodates well the data listed in Table
\ref{tab-nuspeed}.}\label{fig-demo}
\end{figure}

Lacking of a counter example temporarily, let us take a closer look
at this scenario. To get some sense of it, we devised a toy model of
three independent parameters
\begin{equation}\label{toy}
f(\lambda)=\delta\times\exp\left[-\epsilon\left(\frac{\lambda}{\lambda_{\c}}+\frac{\lambda_{\c}}{\lambda}\right)\right]=\delta\times\exp\left[-\epsilon\left(\frac{E_{\c}^2}{E^2}+\frac{E^2}{E_{\c}^2}\right)\right].
\end{equation}
Setting $m_{\nu}=10^{-1}\eV/c^2$, $\epsilon=0.01$,
$\delta=5\times10^{-5}$ and $E_{\c\nu}=5\GeV$, we plot in Fig.
\ref{fig-demo} the dependence of $(v-c)/c$ on $E$. It agrees well
with the observational constraints in Table \ref{tab-nuspeed}. As
illustrated in Fig. \ref{fig-demo}, parameters $E_{\c\nu}$ (or
equivalently $\lambda_{\c}$) and $\epsilon$ dictate the location and
width in energy for Lorentz violation. The magnitude of $f(\lambda)$
or $(v-c)/c$ near $E=E_{\c\nu}$ is controlled by $\delta$. Here the
values of parameters are put and tuned by hand. If there are more
data points, we may numerically compute the best-fit values of
parameters. As indicated by the smallness of $\delta$ and
$\lambda_{\c}$, we need a fine-tuning of parameters in this toy
model. In our scenario, the fine-tuning problem arises from the
energy scale and amplitude of Lorentz violation.

The above toy model is helpful phenomenologically but useless in
construction of a fundamental theory. Although Lorentz symmetry is
broken at the critical scale, we expect a new symmetry will emerge
hereabout. The expected new symmetry should be incorporated in the
theory by devising a more elegant function $f(\lambda)$. The
physical origin of democratic models and new symmetry will be
pondered over in Sec. \ref{sec:disc}.

\subsection{Discriminative Models}\label{subsec:disc}
A more flexible possibility is that the deformed mass-energy
relation is not universal but varies from particle to particle. That
is to say, function $f(\lambda)$ involves more parameters such as
couplings, charges, \emph{etc}. In particular, it is possible that
$f_{\nu}(\lambda)>0$ for neutrinos but $f_i(\lambda)\leq0$ for other
species of particles. The subindex implies the species-dependence.
We feel that this class of models have more chances to survive than
democratic models.

\begin{figure}
\begin{center}
\includegraphics[width=0.45\textwidth]{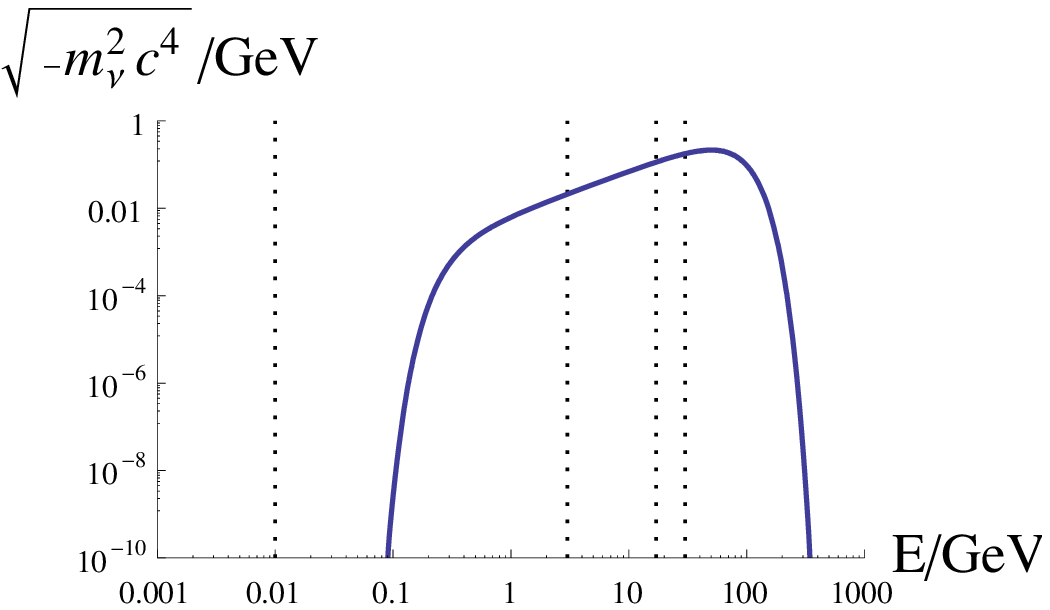}~~
\includegraphics[width=0.45\textwidth]{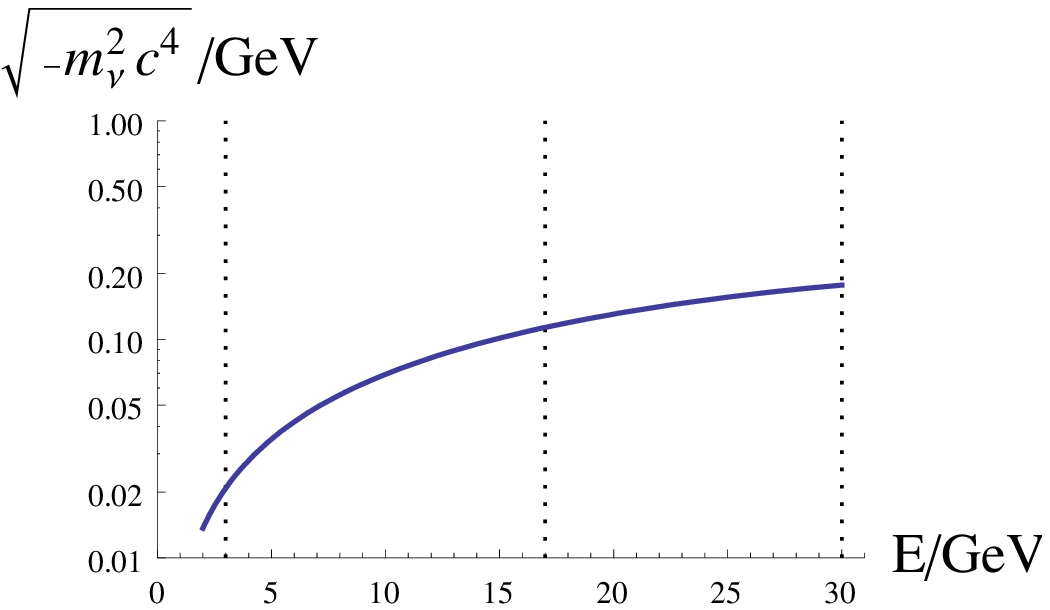}\\
\includegraphics[width=0.45\textwidth]{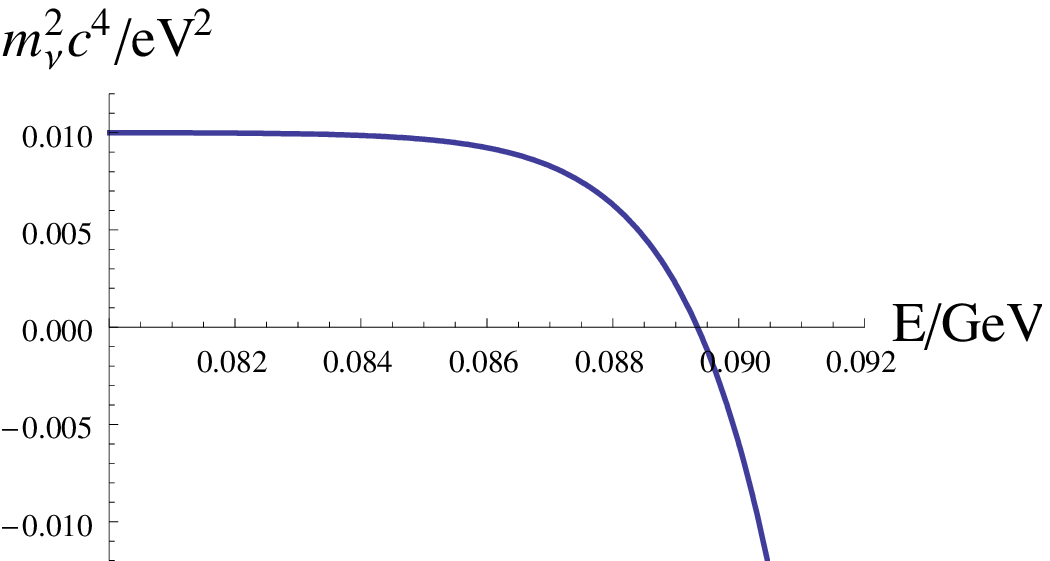}~~
\includegraphics[width=0.45\textwidth]{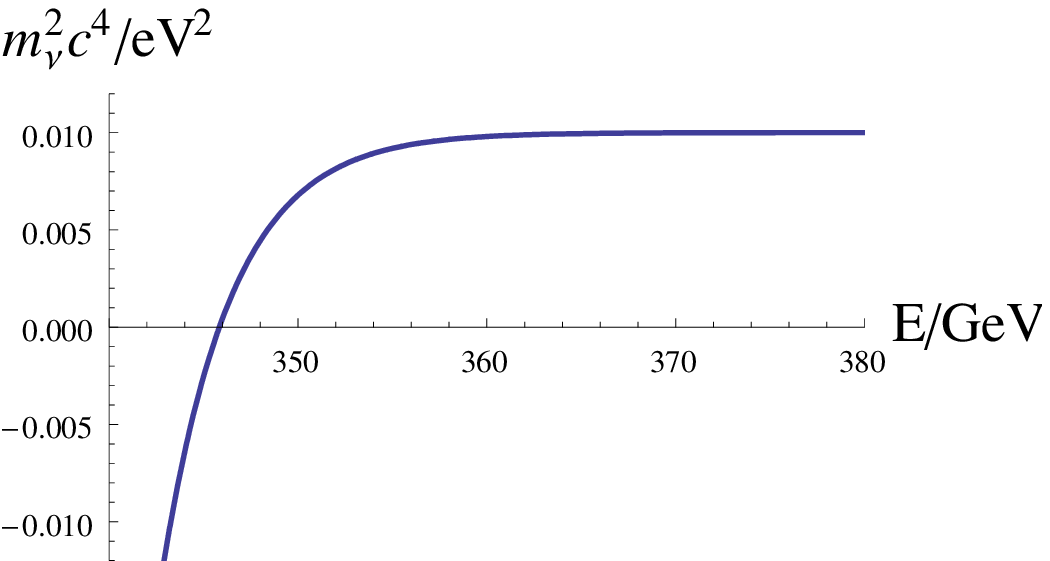}\\
\end{center}
\caption{(color online). The running of neutrino mass at relatively
high energy according to \eqref{toydisc}. The model parameters are
chosen as $m_{\nu}(0)=10^{-1}\eV/c^2$, $\epsilon_{\nu}=0.01$,
$\delta_{\nu}=5\times10^{-5}$ and $E_{\c\nu}=5\GeV$. It is in
agreement with our expectation in Sec.
\ref{sec:tach}.}\label{fig-disc}
\end{figure}

To work with a concrete model, we return to the tachyonic neutrino
with a running mass. Other species are irrelevant in the present
analysis, because $f_i(\lambda)$ for them can be tuned independently
to comply with experiments. Thus we concentrate on the neutrino
sector and phenomenologically apply the toy model \eqref{toy} to it.
After some calculation, we find this can be realized if the
effective squared mass runs like
\begin{equation}\label{toydisc}
m_{\nu}^2(E)=m_{\nu}^2(0)-\frac{E^2}{c^4}f_{\nu}(\lambda)=m_{\nu}^2(0)-\frac{E^2}{c^4}\times\delta_{\nu}\times\exp\left[-\epsilon_{\nu}\left(\frac{E_{\c\nu}^2}{E^2}+\frac{E^2}{E_{\c\nu}^2}\right)\right].
\end{equation}
With parameters similar to that in Fig. \ref{fig-demo}, this model
fit well the given measurement results, as is depicted in Fig.
\ref{fig-disc}. One may compute the running of squared mass for this
model straightforwardly, which changes signature near $E_{c\nu}$. In
model \eqref{toydisc}, $m_{\nu}^2(E)$ crosses zero twice. The
picture of this process was shot and shown in Fig. \ref{fig-disc}.
It indicates that as energy increases, the neutrino runs from an
ordinary particle to a tachyonic particle for a while and then back
to an ordinary particle.

\section{Discussion}\label{sec:disc}
In Sec. \ref{sec:lv}, we accommodated the neutrino velocity anomaly
in the scenario of mass-dependent Lorentz violation. But we did not
explain the physical origins of such a behavior of Lorentz
violation, which remains an open problem. Here are some thoughts on
this problem. As we can expect, there are two possible origins. One
is from the nature of spacetime. This happens in democratic models.
In this case, we would have to ultimately replace the Lorentz
symmetry with some new symmetry, just like switching from Galilean
symmetry to Lorentz symmetry. This will deform plane wave solutions
and would potentially bring troubles to quantization. It reminds us
of doubly special relativity \cite{AmelinoCamelia:2002wr},
$\kappa$-Minkowski spacetime \cite{Agostini:2003dc} and very special
relativity \cite{Cohen:2006ky}. The other origin is from the
dynamics of matter fields, \emph{e.g.} the running of mass with
energy. This happens in both democratic models and discriminative
models. In this case, the Lorentz symmetry is not broken for
spacetime, but it is apparently broken for certain species of matter
at special energy scales.

The results in this paper are tentative rather than conclusive.
First, our investigation has not exhausted all potential
explanations for the superluminal propagation of neutrinos. There
could be reasonable explanations making use of flavor dependence or
time/century/baseline dependence of the data. They are out of the
scope this work although we have highlighted these features in Sec.
\ref{sec:record}. Second, as we emphasized in the very beginning,
before drawing a conclusion about fundamental theories, the most
imperative thing is to further check the systematic errors in
measurement. We are waiting for future news from experiments before
polishing our models.

When this work was near completion, related papers
\cite{Cacciapaglia:2011ax,AmelinoCamelia:2011dx} appeared, also
tackling with the issues of neutrino velocity data in Table
\ref{tab-nuspeed}. Previous investigations related to neutrino
velocity and Lorentz violation can be found from
\cite{Ellis:2008fc,Grossman:2005ej,ThierryMieg:1987nn,Kostelecky:2003cr,Colladay:1998fq,Kostelecky:2008ts,Diaz:2011tx,Pas:2005rb,Dent:2007rk,Hollenberg:2009ws,Alfaro:2004aa}
and references therein.

\begin{acknowledgments}
We are grateful to Xiao-Dong Li and Shuang Wang for valuable
discussion. TW would like to thank the Kavli Institute for
Theoretical Physics China for hospitality where this work was done
during a program on String Phenomenology and Cosmology. ML is
supported by the NSFC grants No.10535060, No.10975172 and
No.10821504, and by the 973 program grant No.2007CB815401 of the
Ministry of Science and Technology of China. TW is supported by the
NSFC grant No.11105053.
\end{acknowledgments}


\begin{thebibliography}{99}
\bibitem{OPERA:2011zb}
  OPERA,
  arXiv:1109.4897 [hep-ex].

\bibitem{Adamson:2007zzb}
  P.~Adamson {\it et al.}  [MINOS Collaboration],
  Phys.\ Rev.\  D {\bf 76}, 072005 (2007)
  [arXiv:0706.0437 [hep-ex]].

\bibitem{Kalbfleisch:1979rm}
  G.~R.~Kalbfleisch, N.~Baggett, E.~C.~Fowler and J.~Alspector,
  Phys.\ Rev.\ Lett.\  {\bf 43}, 1361 (1979).

\bibitem{Alspector:1976kd}
  J.~Alspector {\it et al.},
  Phys.\ Rev.\ Lett.\  {\bf 36}, 837 (1976).

\bibitem{Longo:1987ub}
  M.~J.~Longo,
  Phys.\ Rev.\  D {\bf 36}, 3276 (1987).

\bibitem{Hirata:1987hu}
  K.~Hirata {\it et al.}  [KAMIOKANDE-II Collaboration],
  Phys.\ Rev.\ Lett.\  {\bf 58}, 1490 (1987).

\bibitem{Bionta:1987qt}
  R.~M.~Bionta {\it et al.},
  Phys.\ Rev.\ Lett.\  {\bf 58}, 1494 (1987).

\bibitem{Ciborowski:1996qg}
  J.~Ciborowski and J.~Rembielinski,
  arXiv:hep-ph/9607477.

\bibitem{AmelinoCamelia:2002wr}
  G.~Amelino-Camelia,
  Nature {\bf 418}, 34 (2002)
  [arXiv:gr-qc/0207049].

\bibitem{Agostini:2003dc}
  A.~Agostini,
  arXiv:hep-th/0312305.

\bibitem{Cohen:2006ky}
  A.~G.~Cohen and S.~L.~Glashow,
  Phys.\ Rev.\ Lett.\  {\bf 97}, 021601 (2006)
  [arXiv:hep-ph/0601236].

\bibitem{Cacciapaglia:2011ax}
  G.~Cacciapaglia, A.~Deandrea and L.~Panizzi,
  arXiv:1109.4980 [hep-ph].

\bibitem{AmelinoCamelia:2011dx}
  G.~Amelino-Camelia, G.~Gubitosi, N.~Loret, F.~Mercati, G.~Rosati and P.~Lipari,
  arXiv:1109.5172 [hep-ph].

\bibitem{Ellis:2008fc}
 J.~R.~Ellis, N.~Harries, A.~Meregaglia, A.~Rubbia, A.~Sakharov,
 Phys.\ Rev.\  {\bf D78 } (2008)  033013.
 [arXiv:0805.0253 [hep-ph]].

\bibitem{Grossman:2005ej}
  Y.~Grossman, C.~Kilic, J.~Thaler and D.~G.~E.~Walker,
  Phys.\ Rev.\ D {\bf 72}, 125001 (2005)
  [hep-ph/0506216].

\bibitem{ThierryMieg:1987nn}
  J.~Thierry-Mieg,

\bibitem{Kostelecky:2003cr}
  V.~A.~Kostelecky and M.~Mewes,
  Phys.\ Rev.\  D {\bf 69}, 016005 (2004)
  [arXiv:hep-ph/0309025].

\bibitem{Colladay:1998fq}
  D.~Colladay and V.~A.~Kostelecky,
  Phys.\ Rev.\  D {\bf 58}, 116002 (1998)
  [arXiv:hep-ph/9809521].

\bibitem{Kostelecky:2008ts}
  V.~A.~Kostelecky and N.~Russell,
  Rev.\ Mod.\ Phys.\  {\bf 83}, 11 (2011)
  [arXiv:0801.0287 [hep-ph]].

\bibitem{Diaz:2011tx}
  J.~S.~Diaz,
  arXiv:1109.4620 [hep-ph].

\bibitem{Pas:2005rb}
  H.~Pas, S.~Pakvasa and T.~J.~Weiler,
  Phys.\ Rev.\  D {\bf 72}, 095017 (2005)
  [arXiv:hep-ph/0504096].

\bibitem{Dent:2007rk}
  J.~Dent, H.~Pas, S.~Pakvasa and T.~J.~Weiler,
  arXiv:0710.2524 [hep-ph].

\bibitem{Hollenberg:2009ws}
  S.~Hollenberg, O.~Micu, H.~Pas and T.~J.~Weiler,
  Phys.\ Rev.\  D {\bf 80}, 093005 (2009)
  [arXiv:0906.0150 [hep-ph]].

\bibitem{Alfaro:2004aa}
  J.~Alfaro,
  Phys.\ Rev.\ Lett.\  {\bf 94}, 221302 (2005)
  [arXiv:hep-th/0412295].

\end{thebibliography}
\end{document}